\documentclass[aps,prd,groupedaddress,showpacs,showkeys]{revtex4}
\usepackage{latexsym,epsfig,amsmath,amssymb}
\usepackage{amsfonts}
\usepackage{epsfig} 
\usepackage{verbatim} 
\newcommand{\Jpsi}{J/\psi}

\newcommand{\eq}{\begin{eqnarray}}
\newcommand{\en}{\end{eqnarray}}

\newcommand{\ra}{\rangle}
\begin{document}
\title{Hadronic and radiative three--body decays of $J/\psi$ \\ 
       involving the
       scalars $f_0 (1370)$, $f_0 (1500)$ and $f_0 (1710)$} 
\author{
        Paulos Chatzis, 
        Amand  Faessler,  
        Thomas Gutsche, 
        Valery E. Lyubovitskij\footnote{On leave of absence
        from Department of Physics, Tomsk State University,
        634050 Tomsk, Russia}}
\vspace*{1.2\baselineskip}
\affiliation{
Institut f\"ur Theoretische Physik,  Universit\"at T\"ubingen,\\
Kepler Center for Astro and Particle Physics, \\
Auf der Morgenstelle 14, D--72076 T\"ubingen, Germany\\}
\date{\today}

\begin{abstract}
We study the role of the scalar resonances $f_0(1370)$, $f_0(1500)$ 
and $f_0(1710)$ in the strong and radiative three--body decays of 
$J/\psi$ with $J/\psi \to V + P P (\gamma \gamma)$ and  
$J/\psi \to \gamma + P P (V V)$, where $P (V)$ denotes 
a pseudoscalar (vector) meson. 
We assume that the scalars result from
a glueball-quarkonium mixing scheme while 
the dynamics of the transition process is
described in an effective chiral Lagrangian 
approach. Present data on  $J/\psi \to V + P P$ are well reproduced,
predictions for the radiative processes serve as further tests 
of this scenario.
\end{abstract}


\maketitle

\section{Introduction}

There have been suggestions that mass, production and decay properties
of some of the isoscalar scalar mesons are consistent with a situation 
where a glueball should have mixing with quarkonia states in a mass range
roughly below 2 GeV.
The idea that a glueball configuration is present in the scalar meson
spectrum was suggested in~\cite{Amsler:1995tu} and later extensively used as for example 
in Refs.~\cite{StrohmeierPresicek:1998fi,Close:2000yk,Close:2005vf,%
Zhao:2005ip,Giacosa:2005qr}. 
The discussion on possible evidence for the scalar
ground state glueball has dominantly centered on the scalar-isoscalar
resonances $f_0 (1500)$ and $f_0 (1710)$. The claim that at least partial
glueball nature can be attributed to these states is to some part based
on the mass predictions of Lattice QCD. In the quenched approximation the
lightest glueball state is predicted~\cite{LATTICE1B} to be a scalar with 
a mass of about 1650 MeV and uncertainty of around 100 MeV. 
But note that first results in an unquenched calculation could indicate a strong
downward mass shift towards 1 GeV~\cite{Hart:2006ps}. 
For reviews on the experimental situation of scalar mesons and their 
possible structure interpretation see for example~\cite{Amsler:2004ps}.

An important object for studying the nature of the scalar resonances 
above 1 GeV has been the strong decay patterns observed for these states. 
Thereby several analyses conclude that the three scalar states 
$f_0 (1370)$, $f_0(1500)$ and $f_0 (1710)$ result from a mixing of the 
glueball with the $n \bar n = \sqrt{1/2} \, (u\bar u + d\bar d)$
and $s\bar s$ states of the scalar $^3P_0$ quarkonium nonet. 
Furthermore, the nonappearance or the strength of the appearance of these
scalar states in $\gamma \gamma$ or in central $pp$ collisions 
have been argued to give further signals for a possible quantification of 
the glueball component residing in the 
scalars~\cite{StrohmeierPresicek:1998fi,Close:2000yk,Close:2005vf,%
Zhao:2005ip,Giacosa:2005qr}. 
A study of the strong and radiative three--body decays of $J/\psi$ with 
$J/\psi \to V + P P (\gamma \gamma)$ and  
$J/\psi \to \gamma + P P (V V)$, where 
$V$ and $P$ are vector and pseudoscalar mesons, opens 
a further opportunity to test the nature of the scalar mesons $f_0(1370)$, 
$f_0(1500)$ and $f_0(1710)$. The strong three-body $J/\psi$ decays with 
$PP= \pi^+ \pi^-$ and $K^+ K^-$ in the final state have been studied by 
the Mark III~\cite{Kopke:1987ka}, DM2~\cite{Falvard:1988fc} 
and BES II~\cite{Ablikim:2004st, Ablikim:2004wn} collaborations. 
Particularly the BES II~\cite{Ablikim:2004wn} experiment 
with a much larger statistics indicated a clear signal for 
$f_0(1370) \to \pi^+ \pi^-$ in the $\phi \pi^+ \pi^-$ data.
An enhancement in the $\phi K \bar K$ data can be fitted by
interference between the $f_0(1500)$ and the $f_0(1710)$. 
In~\cite{Bai:2003ww,Ablikim:2006db,Ablikim:2006ca} BES II also reported 
on partial wave analyses on radiative decays of the type 
$J/\psi \to \gamma P P$ and $J/\psi \to \gamma V V$. 
In particular, they indicated the branching ratios 
$B(J/\psi \to\gamma f_0(1710) \to \gamma K \bar K)$~\cite{Bai:2003ww}, 
$B(J/\psi \to\gamma f_0(1710) \to \gamma \pi^+ \pi^-)$, 
$B(J/\psi \to\gamma f_0(1500) \to \gamma \pi^+ \pi^-)$~\cite{Ablikim:2006db} 
and $B(J/\psi \to\gamma f_0(1710) \to 
\gamma \omega \omega)$~\cite{Ablikim:2006ca}. 
 
Theoretical studies of $J/\psi$ decay into a vector meson and two 
pseudoscalars have already been done in 
Refs.~\cite{Morgan:1993rn,Meissner:2000bc,Liu:2009ub}. 
The impact of the effects of coupled--channel dynamics has been 
investigated in~\cite{Morgan:1993rn} and has then been improved on the basis 
of chiral perturbation theory (ChPT) and unitarity constraints 
in~\cite{Meissner:2000bc}. A coherent study of the production of 
the mesons $f_i = f_0(1370)$, $f_0(1500)$ and $f_0(1710)$ in
$J \to V f_i \to V P P$ ($V = \phi, \omega$) has 
been considered in~\cite{Close:2005vf} in the framework based on 
a glueball--quarkonia mixing scheme involving the scalar mesons $f_i$.
An analysis of the role of the scalar mesons $f_0(1370)$ and $f_0(1710)$ 
in $J/\psi$ radiative decays has been performed in 
Ref.~\cite{Zhao:2006dv,Geng:2009iw}. 

It was shown in Refs.~\cite{Korner:1982vg,Kopke:1988cs} that 
the dominant contribution to the Okubo-Zweig-Iizuka (OZI) suppressed 
radiative decays of the $J/\psi$ comes from the photon emission from 
the initial state. This means that the strong $J/\psi$ decays involving
isoscalar vector mesons (like $\omega$ and $\phi$) are not related to the
radiative decays of the $J/\psi$ via the vector meson dominance (VMD) 
mechanism. For this reason the branching ratios of strong and radiative $J/\psi $
decays are compatible.
The three-body $J/\psi$ decays
can be factorized~\cite{Close:2005vf} in terms of the two-body rates. 
Therefore, the analysis of the $J/\psi$ three-body decays can shed 
light on the two-body transitions of the $J/\psi$ and $f_i$ states: 
$J/\psi \to f_i V$, $J/\psi \to f_i \gamma$, 
$f_i \to V \gamma$ and $f_i \to \gamma\gamma$. The electromagnetic 
decays of the scalar mesons $f_i \to V \gamma$ and $f_i \to \gamma\gamma$ 
have been studied in detail using chiral approaches, nonrelativistic 
and light-front quark models (see e.g. 
Refs.~\cite{Giacosa:2005qr,NRQM,LFQM,Nagahiro:2008bn,Nagahiro:2008um,%
Branz:2009cv}).  

The purpose of the present paper is to analyze the role of the scalar mesons 
$f_i$ in the strong and radiative three--body decays of $J/\psi$  
using a chiral Lagrangian approach suggested and developed 
in~\cite{Giacosa:2005qr,Giacosa:2005bw,Gutsche:2008qq}. In these works 
we originally studied the strong and electromagnetic decay properties of 
scalar mesons above 1 GeV. The isoscalar scalars were treated as mixed states
of glueball and quarkonia configurations. Later on we extended the 
phenomenology to the study of decay properties of excited tensor, vector 
and pseudoscalar mesons. Present evaluation is meant as a continuation of 
the full decay analysis presented in~\cite{Giacosa:2005qr} but now we include 
the additional constraints set by $J/\psi$ decays involving these scalars. 
At this stage we do not consider explicitly the mixing of the scalar 
resonances induced by the continuum decay channels, we also neglect final state interaction
in the decay modes (note that inclusion of final-state interaction in $J/\psi$ three-body
decays has been done in~\cite{Liu:2009ub})-- both interaction mechanisms are
higher-order effects in the framework of 
our perturbative considerations.
We are interested to give predictions for scalar resonance contributions to
$J/\psi$ decay modes
where the physical nature of the scalar resonances 
$f_i = f_0(1370)$, $f_0(1500)$ and $f_0(1710)$  
can possibly be tested.

In the present paper we proceed as follows. In Sec.~II, 
we present the effective Lagrangian which will be used for the calculation 
of the resonance contributions of scalar mesons $f_i$ to the matrix elements 
of strong and radiative three--body decays of the $J/\psi$. 
In Sec.~III we present our results for the resonance contributions of 
the scalar mesons $f_i$ to the decay widths of the $J/\psi$. 
A short summary is given in Sec.~IV. 

\section{Approach}
\label{sec:approah} 

In Refs.~\cite{Giacosa:2005qr,Giacosa:2005bw,Gutsche:2008qq} 
we presented the lowest--order chiral Lagrangian describing strong and 
radiative decays of pseudoscalar, scalar, vector and tensor mesons. 
This Lagrangian was motivated by chiral perturbation 
theory~\cite{Weinberg:1978kz,Gasser:1983yg,Gasser:1984gg,Ecker:1988te}. 
Here we extend the formalism by including the additional couplings  
$J/\psi f_i V$, $J/\psi f_i \gamma$, $f_i V V$ and $f_i V \gamma$.
These interaction terms are necessary for elaborating 
the contributions of the scalar mesons $f_i$ to the strong and radiative 
three--body $J/\psi$ decays.  
The full Lagrangian relevant for the $J/\psi$ meson decays involves 
the $J/\psi$ meson, the scalar glueball $G$, 
the nonets of pseudoscalar ${\cal P}$, scalar ${\cal S}$ and vector 
${\cal V}$ mesons with 
\eq 
\label{L_full}
  \mathcal{L}_{\rm eff} 
  &=& \frac{F^2}{4} \langle D_{\mu}UD^{\mu}U^{\dagger} + \chi_+\rangle
  + \frac{1}{2} \langle D_{\mu}{\cal S} D^{\mu} {\cal S} 
  - M^2_S {\cal S}^2\rangle
  + \frac{1}{2} ( \partial_{\mu} G \partial^{\mu} G - M_G^2G^2) 
  \nonumber\\   
  &-& \frac{1}{2}  \langle \nabla_{\mu}  W^{\mu\nu} 
                    \nabla^{\rho} W_{\rho\nu} 
                  - \frac{1}{2} M_{\cal V}^2 W_{\mu\nu} W^{\mu\nu}  \rangle  
   - \frac{1}{2}  (\partial_{\mu}  J^{\mu\nu} 
                  \partial^{\rho} J_{\rho\nu} 
                - \frac{1}{2} M_{J}^2 J_{\mu\nu} 
                    J^{\mu\nu})   
  \nonumber\\   
  &+&c^s_d\langle {\cal S} u_{\mu}u^{\mu}\rangle 
   + c^s_m\langle {\cal S} \chi_+\rangle
   + \frac{c^g_d}{\sqrt{3}}G\langle u_{\mu}u^{\mu}\rangle 
   + \frac{c^g_m}{\sqrt{3}}G\langle \chi_+\rangle
   + \mathcal{L}^{{\cal P}}_{mix} + \mathcal{L}^{{\cal S}}_{mix}
  \nonumber\\
  &+&c^s_e\langle {\cal S} F^+_{\mu\nu}F^{+\mu\nu}\rangle 
   + \frac{c^g_e}{\sqrt{3}}G\langle F^+_{\mu\nu}F^{+\mu\nu}\rangle
   + \frac{c^s_f}{2} \langle {\cal S} \{ W^{\mu\nu}, F^+_{\mu\nu} \} \rangle 
   + \frac{c^g_f}{\sqrt{3}} G \langle W^{\mu\nu} F^+_{\mu\nu} \rangle \\ 
  &+& c^s_v   \langle {\cal S}  W_{\mu\nu} W^{\mu\nu} \rangle 
   + \frac{c^g_v}{\sqrt{3}} G \langle W_{\mu\nu} W^{\mu\nu} \rangle 
   +c^s_h J^{\mu\nu} \langle {\cal S} W_{\mu\nu} \rangle 
   + c^s_w J^{\mu\nu} \langle {\cal S} \rangle \langle W_{\mu\nu} \rangle 
  \nonumber\\
  &+& \frac{c^g_h}{\sqrt{3}} J^{\mu\nu} G \langle W_{\mu\nu} \rangle 
   + J^{\mu\nu} F_{\mu\nu} \, (c_k^n N + c_k^s S + c_k^g G) \, . \nonumber
\end{eqnarray}
The symbols $\langle .. \rangle$ and $\{ \,\, \}$
occurring in Eq.~(\ref{L_full}) denote the trace over flavor 
matrices and the anticommutator, respectively. 
$|N \ra = |(\bar u u + \bar d d)/\sqrt{2} \ra$,  
$|S \ra = | \bar s s \ra$, $|G \ra$ are the nonstrange, strange quarkonia 
states and the glueball, respectively~\cite{Giacosa:2005qr}.  
The constants $c^i_j$ describe the couplings of scalar fields and of 
the glueball to pseudoscalar, vector mesons, $J/\psi$ and to photons. 
For the vector mesons 
we use the tensorial representation in terms of antisymmetric tensor 
fields $W_{\mu\nu}$ (SU(3) nonet of vector mesons) and $J_{\mu\nu}$ ($J/\psi$ 
meson) [see details in~\cite{Ecker:1988te,Kubis:2000zd}]. 
The couplings of the scalars to vector mesons and photons
($c^s_e$, $c^g_e$, $c^s_f$, $c^g_f$, $c^s_v$ and $c^g_v$) 
can be constrained by vector meson 
dominance (VMD) and SU(3) flavor symmetry as: 
\eq\label{VMD_relations} 
c^s_f  &=& 2 \sqrt{2} g_{\rho\gamma} \, c^s_v\,, \quad 
c^g_f \ = \ 2 \sqrt{2} g_{\rho\gamma} \, c^g_v\,, \nonumber\\ 
c^s_e &=& \frac{g_{\rho\gamma}^2}{2} \, c^s_v\,, \quad\quad \hspace*{.25cm} 
c^g_e \ = \  \frac{g_{\rho\gamma}^2}{\sqrt{6}} \, c^g_v \,.  
\en 
As we already stressed in the introduction, the couplings of $J/\psi$ to scalars and photons
contained in $c_k^i$ $ (i=n,s,g)$ are not constrained 
by VMD and, therefore, are independent on the couplings $c_h^s$ 
and $c_h^g$. 
In the derivation of the expressions in Eq.~(\ref{VMD_relations})  
we also used SU(3) relations for the vector--meson to photon couplings: 
\eq\label{Vg_relations}  
g_{\rho\gamma} = 3 g_{\omega\gamma} = \frac{3}{\sqrt{2}} g_{\phi\gamma}
\en 
with $g_{\rho\gamma} = 0.2$ fixed from data on 
$\Gamma(\rho^0 \to e^+ e^-)$~\cite{PDG}.  
The other notations are standard for the basic blocks of 
the ChPT 
Lagrangian~\cite{Weinberg:1978kz,Gasser:1983yg,Gasser:1984gg,Ecker:1988te}: 
$U=u^2=\exp(i P\sqrt{2}/F)$ 
is the chiral field collecting pseudoscalar fields in the exponential 
parametrization, $F$ is the pseudoscalar meson decay constant, 
$D_{\mu}$ and $\nabla$ denote the chiral and gauge-invariant 
derivatives acting on the chiral fields and other mesons. Furthermore, 
$\chi_{\pm} = u^{\dagger}\chi u^{\dagger} \pm u\chi^{\dagger}u$ 
with $\chi = 2B(s+\imath p), \; s = \mathcal{M} + ...$ 
and
$F^+_{\mu\nu} = u^{\dagger}F_{\mu\nu}Qu + uF_{\mu\nu}Qu^{\dagger}$, 
where $F_{\mu\nu}$ is the stress tensor of the electromagnetic field. 
The charge and the current quark mass matrices are represented by 
$Q =  e \, {\rm diag}(2/3, -1/3, -1/3)$ and 
$M = {\rm diag}(\hat{m}, \hat{m}, m_s)$ where $\hat m = (m_u+m_d)/2$ 
and $m_s$ are the nonstrange and strange quark masses.
In the following 
we restrict to the isospin symmetry limit $m_u = m_d$. 
The terms ${\cal L}_{{\rm mix}}^{P}$ and ${\cal L}_{{\rm mix}}^{S}$ 
give rise to flavor singlet-octet mixing in the pseudoscalar and scalar sector,
while for the scalar case quarkonia-glueball mixing is included in addition
[see details in~\cite{Giacosa:2005qr}].
In particular, the mass matrix involving 
nonstrange $|N \ra$ and 
strange $|S \ra = | \bar s s \ra$ quarkonia states and the glueball 
$|G \ra$ has the form~\cite{Giacosa:2005qr}
\eq 
M^2_{\rm bare}=\left( 
\begin{array}{lll}
M_{N}^{2}   & f \sqrt{2} & \varepsilon \\ 
f \sqrt{2}  & M_{G}^{2}  & f \\ 
\varepsilon & f          & M_{S}^{2}
\end{array}
\right) \,,  
\en 
where $f$ and $\varepsilon$ are free parameters controlling 
the quarkonia--glueball mixing scheme. The orthogonal physical states
are obtained by diagonalization of $M^2_{\rm bare}$ with the transformation
matrix $B$ as 
\eq\label{BTr}
B M^2_{\rm bare} B^{T}= M_f^{2} = 
{\rm diag}(M_{f_{1}}^{2}, M_{f_{2}}^{2}, M_{f_{3}}^{2}) \,.
\en 
The eigenvalues of $M_{f}^{2}$ represent the masses of the physical
states $f_{1}\equiv f_{0}(1370)\,,$ $f_{2}\equiv f_{0}(1500)$ and 
$f_{3}~\equiv~f_{0}(1710)\,.$ 
These diagonal states $\left\vert i\right\rangle $, 
with $i=f_{1},f_{2},f_{3}$, are then given in terms of the bare states as 
\eq\label{trafo} 
  \begin{pmatrix}
    |f_1\rangle  \\
    |f_2\rangle \\
    |f_3\rangle
  \end{pmatrix} = 
  B 
  \begin{pmatrix}
    |N\rangle \\
    |G\rangle \\
    |S\rangle 
  \end{pmatrix} \; .
\en

The $\Jpsi$ three--body decays which proceed through the scalar 
resonances are described by the diagrams shown in Figs.1-4. 
The diagram in Fig.1 
contributes to the strong decay $J/\psi \to V f_i \to V P P$. 
There are two graphs in Fig.2 relevant for the radiative decay  
$J/\psi \to V \gamma \gamma$: one described by the two--step process
$J/\psi \to f_i V \to V \gamma \gamma$ [Fig.2(a)] and the other one
with $J/\psi \to f_i \gamma \to V \gamma \gamma$ [Fig.2(b)]. 
Fig.3 contains the graph responsible for the radiative decay 
$J/\psi \to f_i \gamma \to \gamma P P$ involving two pseudoscalars 
in the final state.
Finally, the diagrams of
Figs.4(a) and 4(b), representing the two--step transitions 
$J/\psi \to f_i \gamma \to \gamma V V$ and 
$J/\psi \to f_i V \to \gamma V V$, contribute to the decay 
$J/\psi \to \gamma V V$ with two vector mesons in the decay products.  
The corresponding matrix elements and decay widths resulting from these
specific graphs are given in full detail in the Appendix. 
Note, that the three-body decay widths of $J/\psi$ are factorized 
in terms of two-body decays of the subprocesses 
$J/\psi \to f_i V$, $J/\psi \to f_i \gamma$, $f_i \to P P$, 
$f_i \to V V$, $f_i \to V \gamma$ and $f_i \to \gamma \gamma$. 

\section{Results} 

\subsection{Mixing schemes of the scalar mesons 
$f_0(1370)$, $f_0(1500)$ and $f_0(1710)$} 

In Ref.~\cite{Giacosa:2005qr} we already presented a detailed analysis 
of strong and radiative decays of the scalar mesons 
$f_i = f_0(1370)$, $f_0(1500)$ and $f_0(1710)$ e.g. considering different 
scenarios of quarkonia--glueball mixing. Bare masses, mixing parameters 
and decay couplings of the quarkonia and glueball configurations were 
extracted from a fit to experimental masses
and well-established decay rates into two pseudoscalar mesons.
In the present discussion we also indicate a general fit of these 
parameters, especially
for the decay constants $c^g_d$ and $c^g_m$ of the direct glueball decay.
In Ref.~\cite{Giacosa:2005qr} we constrained these parameters by either 
studying the flavor-symmetry
limit -- $c^g_m=0$ or by matching these constants to first lattice results 
for the glueball
decay. In the following fits no constraints are set on these decay 
parameters, but
previously deduced mixing schemes will approximately result again 
from this procedure. 
The quantities entering in the fitting procedure and the results  
obtained from both fits are displayed in Table I. 
In the fit here we also include the additional 
decay mode $f_0(1500)\to\eta\eta'$. As in~\cite{Giacosa:2005qr} 
$(\Gamma _{f_3})_{2P}$ is the sum of partial decay widths into 
two pseudoscalar mesons approximated by the total width. 
By releasing the constraints on the direct
glueball decay we obtain two solutions where the quality of fits in terms of
$\chi^2_{\rm tot}$ is somewhat improved compared to the ones of 
Ref.~\cite{Giacosa:2005qr}.
We consider two options resulting from local minima in $\chi^2_{\rm tot}$:
Scenario I, the mass of the bare glueball $G$ is on the lower end of 
the unquenched lattice predictions with $m_G \sim 1.5$ GeV; Scenario II, 
the mass of the bare glueball $G$ is larger with $m_G \sim 1.7$ GeV.  
The fit parameters for these two schemes are fixed as follows:  

\noindent 
Scenario I: 
\eq\label{paraI} 
  \begin{tabular}{llrl l llrl l llrl}
    $M_N$      &=&   1.485  & GeV   &,& 
    $M_G$      &=&   1.482  & GeV   &,& 
    $M_S$      &=&   1.698  & GeV \\
    $f$        &=&   0.068  & GeV$^2$ &,& 
    $\epsilon$ &=&   0.236  & GeV$^2$ &,& 
    $c_d^s$    &=&   8.8    & MeV  \\
    $c_m^s$    &=&   2.2    & MeV &,&  
    $c_d^g$    &=&   1.8    & MeV &,& 
    $c_m^g$    &=&  27.7    & MeV  \\
  \end{tabular}
\en 
Scenario II: 
\eq\label{paraII} 
  \begin{tabular}{llrl l llrl l llrl}
    $M_N$      &=&   1.360  & GeV   &,& 
    $M_G$      &=&   1.686  & GeV   &,& 
    $M_S$      &=&   1.439  & GeV \\
    $f$        &=&   0.23   & GeV$^2$ &,& 
    $\epsilon$ &=&   0.30   & GeV$^2$ &,& 
    $c_d^s$    &=&   6.5   & MeV  \\
    $c_m^s$    &=&   5.5   & MeV &,&  
    $c_d^g$    &=&   $-$ 2.0   & MeV &,& 
    $c_m^g$    &=&   48.3    & MeV  \\
  \end{tabular}
\en 
The corresponding mixing matrices $B_{\rm I}$ (Scenario I) 
and $B_{\rm II}$ (Scenario II) are: 
\eq\label{mixI} 
  B_{\rm I} = \begin{pmatrix}
      0.75  &  0.6 &   0.26 \\
     -0.59 &  0.8 &  -0.14 \\
     -0.29 &  -0.05 &   0.95
  \end{pmatrix} \; , 
  \label{RES_MIX1a}
\en
\eq\label{mixII} 
  B_{\rm II} = \begin{pmatrix}
    0.79 &  0.29    & 0.55 \\
    0.57 &   \sim 0 & -0.82 \\
    0.23 &   -0.96 &  0.17
  \end{pmatrix} \; .
  \label{RES_MIX1b}
\en 
For solution I the glueball component dominantly resides in the $f_0 (1500)$
(with $\chi^2_{\rm tot}\simeq 20$), while for the second one 
(with $\chi^2_{\rm tot}\simeq 15$) $f_0 (1710)$ 
is identified with the glueball. Both level schemes of the bare states 
and mixing scenarios contained
in $B_{I,II}$  are very similar to the original solutions
of Ref.~\cite{Giacosa:2005qr}, where the direct glueball decay is included
(denoted in \cite{Giacosa:2005qr} as third and fourth solution). Releasing
the constraint on the direct glueball decay parameters allows a further 
fine-tuning of the results. 
The two solutions also represent the situation discussed
in the literature, when studying the mixing of the glueball with quarkonia
in the  $f_0(1370)$, $f_0(1500)$ and $f_0(1710)$ sector -- the glueball resides
dominantly in the $f_0(1500)$ or the $f_0(1710)$. Qualitatively both
solutions are in acceptable agreement with the experimental data, except for 
the underestimate of $\Gamma_{K_0^*(1430)\to K\pi}$,
which as in the previous work ~\cite{Giacosa:2005qr} remains unresolved. 
In the context of the effective chiral theory the second solution is slightly favored.
We present additional theoretical ratios not contained in Table I, 
compared for example to experimental data from the WA102 
Collaboration~\cite{WA102} 
(see Table II). Note  that in contrast to the result of 
$\Gamma_{f_1\to\bar{K}K}/\Gamma_{f_1\to\pi\pi} = 0.46 \pm 0.19$ 
of the WA102 Collaboration~\cite{WA102} the analysis of the OBELIX 
Collaboration~\cite{Bargiotti:2003ev} results in $0.91 \pm 0.20$. 

\subsection{Hadronic and radiative $\Jpsi$ decays}

In this subsection we analyze the role of the scalar resonances $f_i$ in 
the hadronic $J/\psi$ decays $\Jpsi\to Vf_i\to VPP$. Details of the
calculations are presented in the Appendix: analytical 
formulas for the matrix elements and the decay rates. Our strategy for
determining the parameters involved in these processes is the
following:  using data (central values) on 
the three-body decay modes 
$J/\psi \to (\phi, \omega) f_0 (1710) \to (\phi, \omega) K \bar K$~\cite{PDG} 
(see Table III) as well as the predictions of our approach for the two-body 
decay modes of the scalars (see Table I) we fix the couplings 
$c_h^s$, $c_w^s$, $c_h^g$ in the
effective chiral Lagrangian~(\ref{L_full}) as
\eq 
& &   c_h^{s, \rm I}  = 1.092 \times 10^{-3} \, {\rm GeV}^{-1} \,, 
\quad c_w^{s, \rm I}  = - 0.626 \times 10^{-3} \, {\rm GeV}^{-1} \,, 
\quad c_h^{g, \rm I}  = 10^{-4} \, {\rm GeV}^{-1} \,,  
\nonumber\\ 
& &   c_h^{s, \rm II} = 1.340 \times 10^{-3} \, {\rm GeV}^{-1} \,, 
\quad c_w^{s, \rm II} = - 0.924 \times 10^{-3} \, {\rm GeV}^{-1} \,, 
\quad c_h^{g, \rm II} = 0.722 \times 10^{-3} \, {\rm GeV}^{-1} \,. 
\en  
The indices I, II refer to the respective mixing scenarios. With these
couplings we determine the strong couplings $g_{Jf_iV}$ [they are 
linear combinations of $c_h^s$ and $c_h^g$~(\ref{g_JfV})] and 
the electromagnetic couplings $g_{Jf_i\gamma}$ of the physical 
$f_i$ states. In Table IV we list the results for the effective  
couplings $g_{Jf_iV}$ and $g_{Jf_i\gamma}$ and the corresponding two-body  
decay widths.  

In Table III we give our final results for the hadronic three-body 
decays of $J/\psi$ for both scenarios (I and II). Results are compared
to data~\cite{PDG} and the predictions of Ref.~\cite{Close:2005vf},
where a similar mixing scheme ~\cite{Amsler:1995tu} has been
used to determine $B$. Especially in scenario I we obtain a
reasonable description of present data on the hadronic $J/\psi $ decays,
while in version II some discrepancies occur.   
In particular, one can see that in scenario I  
we can reproduce the central value of the ratio 
\eq 
R = \frac{{\rm Br}(\Jpsi\to f_3\phi \to \phi K \bar K)}
           {{\rm Br}(\Jpsi\to f_3\omega \to \omega K \bar K)} = \frac{4}{3} 
\en 
while in scenario II this quantity is close to 2. The explanation for this quantitative difference
between the two mixing schemes is quite simple. The ratio R is approximately equal to the 
ratio of the corresponding strong two-body decay modes of $J/\psi$, 
which is expressed in terms of the elements of the mixing matrix $B$ as: 
\eq
R \simeq \frac{{\rm Br}(\Jpsi\to f_3\phi)} 
              {{\rm Br}(\Jpsi\to f_3\omega)} = 
2 \biggl[ 1 + \frac{c_h^s (B_{31}/\sqrt{2} - B_{33})}
{c_h^s B_{33} + c_h^g B_{32}/\sqrt{3} + c_w^s (B_{31}\sqrt{2} + B_{33})} 
\biggr]^2 \,. 
\en 
The ratio depends crucialy on the value of the combination of mixing 
matrix elements with $\Delta = B_{31}/\sqrt{2} - B_{33}$. In scenario I
we have $\Delta =-1.16$ and therefore it is possible to generate the
observed value for the 
ratio $R$ with the appropriate choice of free parameters 
$c_h^s$, $c_h^g$ and $c_w^s$ (as we did in our fitting procedure). 
In case of scenario II this quantity has the value $\Delta = - 0.007 \simeq 0$, 
and now the ratio $R$ is always close to $2$ independent on 
the choice of the relevant parameters $c_h^s$, $c_h^g$ and $c_w^s$.
After the mxing matrix $B$ is constrained by a fit to the properties
of the scalars (as compiled in Table I) scenario II is less preferable when
analysing the hadronic  $J/\psi$ decays.

Analysis of the radiative $\Jpsi$ decays 
$J/\psi \to \gamma + P P (V V)$ and $J/\psi \to P + \gamma\gamma$
can produce further important information 
about the electromagnetic structure of the scalar resonances involved.
The relevant matrix elements involve further unknown 
couplings ($c^s_v$, $c^s_f$, $c^s_e$, $c^g_v$, $c^g_f$, $c^g_e$, 
$c^n_k$, $c^s_k$, $c^g_k$) from the effective 
Lagrangian~(\ref{L_full}). Note, only 5 parameters from this set 
are independent, because the parameters $c^s_f$, $c^s_e$ and 
$c^g_f$, $c^g_e$ are expressed through the parameters $c^s_v$ 
and $c^g_v$ via VMD constraints~(\ref{VMD_relations}). 

The other parameters involved were already fixed from the previous analysis 
of the strong $J/\psi$ decays. We find that acceptable results for the 
radiative $J/\psi$ decays can be achieved with the following choice of the 
parameters $c^s_v$, $c^g_v$, $c^n_k$, $c^s_k$, $c^g_k$: 
\eq\label{em_chiral} 
& &   c_v^{s, \rm I}  = 2.815   \ {\rm GeV}^{-1} \,, 
\quad c_v^{g, \rm I}  = 0.138   \ {\rm GeV}^{-1} \,, \nonumber\\
& & c_k^{n, \rm I}    =   0.241   \times 10^{-2} \ {\rm GeV}^{-1} \,, 
\quad c_k^{s, \rm I}  = - 0.313   \times 10^{-2} \ {\rm GeV}^{-1} \,, 
\quad c_k^{g, \rm I}  = - 0.271   \times 10^{-3} \ {\rm GeV}^{-1} \,, 
\nonumber\\ 
& &\\
& &   c_v^{s, \rm II} = 3.036   \ {\rm GeV}^{-1}  \,, 
\quad c_v^{g, \rm II} = 1.898  \ {\rm GeV}^{-1}\,,  \nonumber\\
& & c_k^{n, \rm II}   = 0.132  \times 10^{-2} \ {\rm GeV}^{-1}\,. 
\quad c_k^{s, \rm II} = 0.187  \times 10^{-2} \ {\rm GeV}^{-1} \,, 
\quad c_k^{g, \rm II} = - 0.352  \times 10^{-2} \ {\rm GeV}^{-1} \,.\nonumber
\en    
Knowledge of the couplings $c_v^s$ and $c_v^g$ gives access to the 
effective couplings of scalar mesons to vector mesons and photons. 
They are related by the VMD relations~(\ref{VMD_relations}) and 
(\ref{Vg_relations}). 
With the use of Eq.~(\ref{em_chiral}) and the VMD relations we determine the 
effective couplings of the scalar $f_i$ mesons to photons and vector mesons 
as listed in Table V.

Our predictions for the radiative decays of
$\Jpsi$ and the scalar mesons $f_i$ are given in Tables VI and VII.   
Note that we reproduce all known data for 
radiative decays of the $J/\psi$ involving the $f_i$ states~\cite{PDG}: 
\eq 
& &{\rm Br}(\Jpsi\to f_2\gamma) = (1.01 \pm 0.32) \times 10^{-4}\,, 
\nonumber\\ 
& &{\rm Br}(\Jpsi\to f_3\gamma\to \gamma \pi\pi) 
= (4.0 \pm 1.0) \times 10^{-4}\,, \\
& &{\rm Br}(\Jpsi\to f_3\gamma\to \gamma K\bar{K}) 
= (8.5^{+1.2}_{-0.9}) \times 10^{-4}\,, \nonumber\\
& &{\rm Br}(\Jpsi\to f_3\gamma\to \gamma \omega\omega) 
= (3.1 \pm 1.0) \times 10^{-4}\,. \nonumber
\en 
In case of the scalar mesons we set our results in comparison to the ones
of other approaches and with available data. 
In particular, the results of 
Ref.~\cite{NRQM} and~\cite{LFQM} are given in the form
$(L,M,H)$, where $L$=Light, $M$=Medium and $H$=Heavy correspond to the 
three possibilities for the bare glueball mass: 
lighter than the bare $\bar n n$ mass, 
between the $\bar n n$ and $\bar s s$ masses, heavier than the $\bar s s$ 
mass. The results 
of Ref.~\cite{Nagahiro:2008bn} are listed in the form $(K,P)$, where 
$K=K\bar{K}$ and $P=\pi\pi$ indicate the contributions of  
the intermediate $K\bar{K}$ and $\pi\pi$ loop, respectively.

The calculated branchings of the two- and three-body decays 
of $J/\psi$ and the $f_i$ mesons satisfy the following approximate 
ratios: 
\eq 
R_{i1} &=& \frac{{\rm Br}(\Jpsi\to f_i\gamma\to\pi\pi\gamma)}
              {{\rm Br}(\Jpsi\to f_i\gamma\to K \bar K \gamma)} \, \ \simeq 
        \, \ R_{i2} \,\, \ = \, \ \frac{{\rm Br}(f_i\to\pi\pi)}
{{\rm Br}(f_i\to K \bar K)}\,, \nonumber\\
R_{i3}^{VPP} &=& \frac{{\rm Br}(\Jpsi\to f_i V \to P P V)}
              {{\rm Br}(\Jpsi\to f_i V)} \ \simeq \  
R_{i4}^{PP} \ = \ \frac{{\rm Br}(\Jpsi\to f_i V \to P P \gamma)}
              {{\rm Br}(\Jpsi\to f_i \gamma)} 
        \ \simeq \ {\rm Br}(f_i\to P P) \,. 
\en 
We want to illustrate these
approximate constraints in case of the scenario I.  
The set of the radiative three-body $J/\psi$ branching ratios 
$R_{11} = 1.04$, $R_{21} = 3.80$, 
$R_{31} = 0.56$ is similar to the corresponding set 
of the strong two-body $f_i$ branching ratios 
$R_{12} = 0.94$, $R_{22} = 3.61$, 
$R_{32} = 0.43$. 
Also, the ratios $R_{i3}^{VPP}$ and $R_{i4}^{PP}$ are approximately equal 
to the corresponding branchings of the strong two-body decays of the scalar mesons 
$f_i$. E.g. 
in case of $f_3=f_0(1710)$ and $K\bar K$ in the final state we 
have 
\eq 
R_{33}^{\omega KK} \ = \ 0.45 \ \simeq \ R_{33}^{\phi KK} \ = \ 0.44 
\ \simeq \ 
R_{34}^{KK} \ = \ 0.45 \ \simeq \ {\rm Br}(f_3\to K\bar K) \ = \ 0.49 \,.
\en 
Similar consistency is found in case of the scenario II with: 
\eq 
& &R_{11} = 0.51 \ \simeq \ R_{12} = 0.44\,, \quad 
   R_{21} = 0.41 \ \simeq \ R_{21} = 0.39\,, \quad   
   R_{31} = 0.64 \ \simeq \ R_{31} = 0.49\,, \\
& &R_{33}^{\omega KK} \ \simeq \ R_{33}^{\phi KK} \ \simeq 
   R_{34}^{KK} \ = \ 0.38 \ \simeq \ {\rm Br}(f_3\to K\bar K) \ = \ 0.42 \,.
\en
  
\section{Conclusions}
 
In conclusion, we present an analysis of the 
role of the scalar resonances $f_0(1370)$, $f_0(1500)$ 
and $f_0(1710)$ in the strong and radiative
$J/\psi$ three-body decays. 
This analysis can shed light on the nature of the scalar 
resonances $f_i$ and the possibility that these states
arise from glueball-quarkonia mixing. 
We tested two cases for the glueball-quarkonia mixing schemes: 
scenario I, the bare glueball dominantly resides in the $f_0 (1500)$;
scenario II, the scalar $f_0 (1710)$ contains the largest glueball
component. We found that the first  
scenario is more consistent with present data, especially for the case of the
strong three-body decays of $J/\psi$.  The results are also consistent 
with available data on radiative decays. The detailed set of 
predictions for the radiative
decays can serve to further distinguish between the different structure 
assumptions of the scalars. It would therefore be extremely useful to have 
more data on the radiative processes.

\begin{acknowledgments}

This work was supported by the DFG under Contract No. FA67/31-2. 
This research is also part of the European
Community-Research Infrastructure Integrating Activity
``Study of Strongly Interacting Matter'' (acronym HadronPhysics2,
Grant Agreement No. 227431), Federal Targeted Program "Scientific 
and scientific-pedagogical personnel of innovative Russia" 
Contract No. 02.740.11.0238. 

\end{acknowledgments}

\appendix 
\section{Matrix elements and decay widths of the $\Jpsi$ three-body decays} 

For the calculation of the three--body decays of the $\Jpsi$ we 
introduce several notations --- 
$p$ is the momentum of the $\Jpsi$ and $q$, $q_1$, $q_2$, are 
the momenta of the final state particles: $V(q), P(q_1), P(q_2)$ 
[for the decay $J/\psi \to V f_i \to V P P]$,  
$V(q), \gamma(q_1), \gamma(q_2)$ 
[decay $J/\psi \to f_i V(\gamma) \to V \gamma \gamma]$, 
$\gamma(q), P(q_1), P(q_2)$ [decay $J/\psi \to f_i \gamma \to \gamma P P]$ 
and  
$\gamma(q), V(q_1), V(q_2)$ [decay $J/\psi \to f_i V(\gamma) \to \gamma V V]$. 

We also define the invariant variables $s_i (i=1,2,3)$: 
\eq 
p   &=& q + q_1 + q_2 \,, \nonumber\\
s_1 &=& (q + q_1)^2   = (p - q_2)^2  \,, \nonumber\\
s_2 &=& (q_1 + q_2)^2 = (p - q)^2  \,, \nonumber\\
s_3 &=& (q + q_2)^2   = (p - q_1)^2  \,.  
\en 
The matrix elements describing the $J/\psi$ three-body strong 
and radiative decays proceeding through a scalar resonance $f_i$
are denoted by $M_{\rm I (II, III, IV)}^{a (b)}$ with super- and subscripts 
referring to the graphs of Figs.1-4 in an obvious notation. The explicit 
expressions for these amplitudes read as
\eq 
M_{\rm I} &\equiv& M(J/\psi \to f_i V\to V P P) \ = \ 
\epsilon_J^\mu(p) \, \epsilon_V^{\ast \nu}(q) \ 
g_{Jf_iV} \  g_{f_iPP} \ (g_{\mu\nu} \, pq  - p_{\nu}q_{\mu}) \  
\frac{1}{M_{f_i}^2 - s_2 - i M_{f_i}\Gamma_{f_i}} \,, \\
M_{\rm II} &\equiv& M(J/\psi \to f_i V(\gamma) \to V \gamma \gamma) 
\ = \ M^{(a)}_{\rm II}+M^{(b)}_{\rm II} \,, \\
M^{(a)}_{\rm II} &\equiv& M(J/\psi \to f_i \gamma \to V \gamma\gamma) =  
\epsilon_J^{\mu}(p) \,  \epsilon_V^{\ast \nu}(q) \, 
\epsilon_{\gamma}^{\ast \alpha}(q_1) \,  
\epsilon_{\gamma}^{\ast \beta}(q_2) \ 
g_{Jf_iV} \  g_{f_i\gamma\gamma} \nonumber\\
&\times&  (g_{\mu\nu} \, pq - p_{\nu}q_{\mu}) 
(g_{\alpha\beta}  q_1q_2-q_{1\beta}q_{2\alpha})  
\frac{1}{M_{f_i}^2 - s_2 - i M_{f_i}\Gamma_{f_i}}\,, \nonumber\\
M^{(b)}_{\rm II} &\equiv& M(J/\psi \to f_i V \to V \gamma\gamma) =  
\epsilon_J^{\mu}(p) \, \epsilon_V^{\ast \nu}(q)  
\, \epsilon_{\gamma}^{\ast \alpha}(q_1) 
\, \epsilon_{\gamma}^{\ast \beta}(q_2)  
\ g_{Jf_i\gamma} \ g_{f_i V \gamma} \nonumber\\
&\times& 
\biggl[ (g_{\mu\alpha} pq_1 - p_{\alpha}q_{1\mu})(g_{\nu\beta} qq_2 
- q_{\beta}q_{2\nu}) \frac{1}{M_{f_i}^2 - s_3 - i M_{f_i}\Gamma_{f_i}}
\nonumber\\
&+&        
(g_{\mu\beta}  pq_2 - p_{\beta} q_{2\mu})(g_{\nu\alpha} \, qq_1 
- q_{\alpha}q_{1\nu}) \frac{1}{M_{f_i}^2 - s_1
- i M_{f_i}\Gamma_{f_i}}\biggr] \,, \nonumber\\ 
M_{\rm III} &\equiv& M(J/\psi \to \gamma f_i \to \gamma P P) \ = \ 
\epsilon_J^\mu(p) \, \epsilon_\gamma^{\ast \nu}(q) \ 
g_{Jf_iV} \  g_{f_iPP} \ (g_{\mu\nu} \, pq  - p_{\nu}q_{\mu}) \  
\frac{1}{M_{f_i}^2 - s_2 - i M_{f_i}\Gamma_{f_i}} \,, \\ 
M_{\rm IV} &\equiv& M(J/\psi \to f_i V(\gamma) \to \gamma V V) 
\ = \ M^{(a)}_{\rm IV}+M^{(b)}_{\rm IV} \,, \\
M^{(a)}_{\rm IV} &\equiv& M(J/\psi \to f_i \gamma \to \gamma V V) =  
\epsilon_J^{\mu}(p) \,  \epsilon_\gamma^{\ast \nu}(q) \, 
\epsilon_{V}^{\ast \alpha}(q_1) \,  
\epsilon_{V}^{\ast \beta}(q_2) \ 
g_{Jf_i\gamma} \  g_{f_iVV} \nonumber\\
&\times&  (g_{\mu\nu} \, pq - p_{\nu}q_{\mu}) 
(g_{\alpha\beta}  q_1q_2-q_{1\beta}q_{2\alpha})  
\frac{1}{M_{f_i}^2 - s_2 - i M_{f_i}\Gamma_{f_i}}\,, \nonumber\\
M^{(b)}_{\rm IV} &\equiv& M(J/\psi \to f_i V \to \gamma V V) =  
\epsilon_J^{\mu}(p) \, \epsilon_\gamma^{\ast \nu}(q)  
\, \epsilon_{V}^{\ast \alpha}(q_1) 
\, \epsilon_{V}^{\ast \beta}(q_2)  
\ g_{Jf_iV} \ g_{f_i V \gamma} \nonumber\\
&\times& 
\biggl[ (g_{\mu\alpha} pq_1 - p_{\alpha}q_{1\mu})(g_{\nu\beta} qq_2 
- q_{\beta}q_{2\nu}) \frac{1}{M_{f_i}^2 - s_3 - i M_{f_i}\Gamma_{f_i}}
\nonumber\\
&+&        
(g_{\mu\beta}  pq_2 - p_{\beta} q_{2\mu})(g_{\nu\alpha} \, qq_1 
- q_{\alpha}q_{1\nu}) \frac{1}{M_{f_i}^2 - s_1
- i M_{f_i}\Gamma_{f_i}}\biggr] \,, \nonumber  
\en 
where $M_{f_i}$ and $\Gamma_{f_i}$ are the mass and width of the scalar meson 
$f_i$. Here, the coefficients $g_{Jf_iV}$, $g_{f_iPP}$, 
$g_{f_iVV}$, $g_{Jf_i\gamma}$, $g_{f_iV\gamma}$ and $g_{f_i\gamma\gamma}$ 
are the couplings involving mixed scalars, which are related to the
couplings of the effective Lagrangian~(\ref{L_full}) involving unmixed 
states as:  
\eq 
\hspace*{-.5cm}
g_{J f_i \omega} &=& 2 B_{i1} c^s_h + 2 \sqrt{\frac{2}{3}} B_{i2} c^g_h 
+ 2 \sqrt{2} (B_{i1} \sqrt{2} + B_{i3}) c_w^s\,, 
\hspace*{.15cm} 
g_{J f_i \phi} \ = \ - 2 B_{i3} c^s_h - 2 \sqrt{\frac{1}{3}} B_{i2} c^g_h 
- 2 (B_{i1} \sqrt{2} + B_{i3}) c_w^s\,, \label{g_JfV}\\  
\hspace*{-.5cm}
g_{f_i \pi\pi} &=&
- \frac{2\,B_{i1}}{F^{2}\, 
\sqrt{2}}\,\biggl( (M_{f_i}^{2}\,-\,2\,M_{\pi }^{2}) 
\,c_{d}^{s}\,+\,2\,M_{\pi}^{2}\,c_{m}^{s}\biggr)\,
- \,\frac{2\,B_{i2}}{F^{2}\,\sqrt{3}}
\biggl( (M_{f_i}^{2}\,-\,2\,M_{\pi }^{2})\,c_{d}^{g}\,
+\,2\,M_{\pi }^{2}\,c_{m}^{g} \biggr)\,, \\
\hspace*{-.5cm}
g_{f_i K K} &=& - 
\frac{B_{i1}+\sqrt{2}B_{i3}}{F^{2}\sqrt{2}}
\,\biggl( (M_{f_i}^{2}\,-\,2\,M_{K}^{2})\,c_{d}^{s}\,+\,2 
\,M_{K}^{2}\,c_{m}^{s}\biggr)  
-\frac{2\,B_{i2}}{F^{2}\,\sqrt{3}}\biggl( (M_{f_i}^{2}\,-\,2\,M_{K}^{2})
\,c_{d}^{g}\,+\,2\,M_{K}^{2}\,c_{m}^{g}\biggr)\,, \\
\hspace*{-.5cm}
g_{f_i \rho\rho} &=& g_{f_i \omega\omega} \ = \ 
\frac{4}{\sqrt{2}} c^s_v B_{i1} 
+ \frac{4}{\sqrt{3}} c^g_v B_{i2}\,, \hspace*{.5cm} 
g_{f_i \phi\phi} \ = \ 4 c^s_v B_{i3} 
+ \frac{4}{\sqrt{3}} c^g_v B_{i2} \,, \\ 
\hspace*{-.5cm}
g_{J f_i \gamma} &=& 2 ( B_{i1} c_k^n + B_{i2} c_k^g + B_{i3} c_k^s ) \,, 
\hspace*{.25cm} 
 g_{f_i\gamma\gamma} \ = \ \frac{16}{9} 
\biggl( \frac{5}{\sqrt{2}} B_{i1} + B_{i3} \biggr) c^s_e 
+ \frac{32}{3\sqrt{3}} B_{i2} c^g_e \,, \label{g_figg}\\
\hspace*{-.5cm}
g_{f_i\rho\gamma} &=& 3 g_{f_i\omega\gamma} = 
B_{i1} c^s_f + \sqrt{\frac{2}{3}} B_{i2} c^g_f  \,, \hspace*{.5cm}  
g_{f_i\phi\gamma} \ = \
\frac{2}{3} B_{i3} c^s_f + \frac{2}{3 \sqrt{3}} B_{i2} c^g_f \, .
\label{gfrhog}
\en 
In above expressions the
$B_{ij}$ are the elements of the matrix $B$ relating 
the mixed $(f_1, f_2, f_3)$ and unmixed $(N, G, S)$ scalar states as in
Eq.~(\ref{trafo}).
If the process is kinematically allowed, the 
couplings $g_{Jf_iV}$, $g_{Jf_i\gamma}$, $g_{f_iPP}$, $g_{f_iVV}$, 
$g_{f_iV\gamma}$ and $g_{f_i\gamma\gamma}$ define the corresponding 
two-body decay widths of $J/\psi$ and the physical $f_i$ states as: 
\eq 
\Gamma(\Jpsi \to f_i V) &=& \frac{g_{Jf_iV}^2}{16 \pi M_J} 
 \, M_V^2 \, \lambda^{1/2}(M_J^2,M_{f_i}^2,M_V^2) 
\biggl(1 + \frac{\displaystyle{\lambda(M_J^2,M_{f_i}^2,M_V^2)}}
{\displaystyle{6M_J^2M_V^2}}\biggr)  
\,, \label{gJfV}\\ 
\Gamma(f_i \to PP) &=& \frac{g_{f_iPP}^2}{16 \pi M_{f_i}} \, 
N_{PP} \, \sqrt{1 - \frac{4M_P^2}{M_{f_i}^2}} \label{gfPP}\\ 
\Gamma(f_i \to VV) &=& \frac{g_{f_iVV}^2}{64 \pi} \, 
M_{f_i}^3 \, \sqrt{1 - \frac{4M_V^2}{M_{f_i}^2}} 
\biggl( 1 -  \frac{4M_V^2}{M_{f_i}^2} 
+ \frac{6M_V^4}{M_{f_i}^4} \biggr) \label{gfVV}\\ 
\Gamma(J/\psi \to f_i\gamma) &=& \frac{\alpha}{24} \, 
g_{Jf_i\gamma}^2 \, M_J^3 \, 
\biggl(1 - \frac{M_{f_i}^2}{M_J^2} \biggr)^3 \,, \label{gJfg}\\
\Gamma(f_i\to V\gamma) &=& \frac{\alpha}{8} \, g_{f_iV\gamma}^2 \, 
M_{f_i}^3 \biggl(1- \frac{M_V^2}{M_{f_i}^2}\biggr)^3 \,, 
\label{gfVg}\\ 
\Gamma(f_i \to \gamma\gamma) &=& \frac{\pi}{4} \, 
\alpha^2 \, g_{f_i\gamma\gamma}^2 \, M_{f_i}^3 \,. 
\label{gfgg} 
\en 
where $\lambda(x,y,z) = x^2 + y^2 + z^2 - 2 xy - 2 xz - 2 yz$ 
is the K\"allen function. The factor
\eq  
N_{PP} = 
  \begin{cases}
    \frac{3}{2},  & PP = \pi \pi  \\
    2,            & PP = K \bar K    \\ 
  \end{cases}
\en 
takes into account the sum over charged modes: 
\eq
\Gamma(f_i \to \pi\pi) &=& \Gamma(f_i \to \pi^+\pi^-) 
+ \Gamma(f_i \to \pi^0\pi^0)\,, \nonumber\\
\Gamma(f_i \to K \bar K) &=& \Gamma(f_i \to K^+ K^-) 
+ \Gamma(f_i \to K^0 \bar K^0) \,.  
\en   
Note, the couplings $g_{f_i VV}$, $g_{f_i V\gamma}$,  
$g_{f_i \gamma\gamma}$ and $g_{J f_i V}$, $g_{J f_i \gamma}$  
are constrained by VMD and SU(3) flavor symmetry relations
(see also Eqs.~(\ref{VMD_relations}) and~(\ref{Vg_relations})):     
\eq 
g_{f_i\gamma\gamma} &=& 
g_{\rho\gamma} \biggl( g_{f_i\rho\gamma} + \frac{1}{3} g_{f_i\omega\gamma} 
+ \frac{\sqrt{2}}{3} g_{f_i\phi\gamma} \biggr)  
\ = \ 
g_{\rho\gamma}^2 \biggl( g_{f_i\rho\rho} + \frac{1}{9} g_{f_i\omega\omega} 
+ \frac{2}{9} g_{f_i\phi\phi} \biggr)\,, \label{VMD1}\\ 
g_{f_i \rho \gamma} &=& g_{\rho\gamma} g_{f_i \rho \rho}\,, \quad 
g_{f_i \omega \gamma} \ = \ \frac{1}{3} 
g_{\rho\gamma} g_{f_i \omega \omega}\,, \quad 
g_{f_i \phi \gamma} \ = \ \frac{\sqrt{2}}{3} 
g_{\rho\gamma} g_{f_i \phi \phi}\,, \label{VMD2}\\  
g_{J f_i \gamma} &=& g_{\rho\gamma} \biggl( \frac{1}{3} g_{J f_i \omega} 
+ \frac{\sqrt{2}}{3} g_{J f_i \phi} \biggr) \label{VMD3} 
\en 
with $g_{\rho\gamma} = 0.2$. 

Next we write down the expressions for the $J/\psi$ three-body decay widths.  
The decay width $\Gamma_{\rm I} \equiv \Gamma(\Jpsi \to f_i V \to V P P)$  
is given by: 
\eq 
\Gamma_{\rm I} &=& \frac{N_{PP}}{768 \pi^3 M_J^3}  
\int\limits_{4M_P^2}^{(M_J-M_V)^2}
ds_2 \int\limits_{s_{\rm 1, I}^-}^{s_{\rm 1, I}^+}ds_1 
\sum_{\rm pol}|M_{\rm I}|^2  
= \frac{g_{Jf_iV}^2 \, g_{f_iPP}^2}{1536 \pi^3 \, M_J^3} \ 
N_{PP} \ R \,, 
\en 
where the sum is performed over the polarizations and where  
\eq
R &=& \int\limits_{4M_P^2}^{(M_J-M_V)^2} ds_2  
\frac{\lambda^{1/2}(M_J^2,M_V^2,s_2)} 
{(M_{f_i}^2 - s_2)^2 + \Gamma_{f_i}^2 M_{f_i}^2}  \, 
\sqrt{1-\frac{4M_P^2}{s_2}}
\, \biggl( \lambda(M_J^2,M_V^2,s_2) + 6 M_J^2 M_V^2 \biggr) \,,
\nonumber \\
s_{\rm 1, I}^{\pm} &=& M_P^2 + \frac{1}{2} \biggl( M_J^2 + M_V^2 - s_2 
\pm  \lambda^{1/2}({s_2,M_J^2,M_V^2}) \, 
\sqrt{1-\frac{4M_P^2}{s_2}} \ \biggr) \, . 
\en 
Using Eqs.~(\ref{gJfV}) and (\ref{gfPP}) we express $\Gamma_{\rm I}$ 
through the two-body decay widths $\Gamma(\Jpsi \to f_i V)$ and 
$\Gamma(f_i \to PP)$: 
\eq\label{GammaI_fin} 
\Gamma_{\rm I} &=& \frac{\Gamma(\Jpsi \to f_i V) \, 
\Gamma(f_i \to PP)}{\pi \, \lambda^{1/2}(M_J^2,M_{f_i}^2,M_V^2) \, 
\sqrt{M_{f_i}^2 - 4M_P^2}} \, 
\frac{M_{f_i}^2}{\lambda(M_J^2,M_{f_i}^2,M_V^2) + 6 M_J^2M_V^2} \ R \,. 
\en 

The decay width $\Gamma_{\rm II} \equiv  
\Gamma(J/\psi \to f_i V(\gamma) \to V \gamma\gamma)$ 
is calculated according to the expression: 
\eq 
\label{GammaII}
\Gamma_{\rm II} = \frac{1}{1536 \pi^3 M_J^3} \int\limits_{0}^{(M_J-M_V)^2}ds_2 
\int\limits_{s_{\rm 1, II}^-}^{s_{\rm 1, II}^+} 
ds_1 \sum_{\rm pol}|M_{\rm II}|^2 \, , 
\en 
where 
\eq 
s_{\rm 1, II}^\pm = \frac{M_J^2 + M_V^2 - s_2}{2} \pm 
\frac{\lambda^{1/2}(s_2,M_J^2,M_V^2)}{2} \, .
\en
The sum over the polarizations is rewritten as
\eq
\sum_{\rm pol}|M_{\rm II}|^2 
= 4 \pi^2 \alpha^2 \biggl( 
g_{Jf_iV}^2 \ g_{f_i\gamma\gamma}^2 \ R_1 
 + g_{Jf_i\gamma}^2 \ g_{f_iV\gamma}^2 \ R_2
 + g_{Jf_iV} \ g_{Jf_i\gamma} \ g_{f_iV\gamma} \ g_{f_i\gamma\gamma} 
\ R_3 \biggr) 
\en 
where
\eq 
R_1 &=& \frac{s_2^2}{(M_{f_i}^2 - s_2)^2 + M_{f_i}^2 \Gamma_{f_i}^2} \ 
\biggl(\lambda(M_J^2,M_V^2,s_2)+6 M_J^2 M_V^2\biggr)\,, \\ 
R_2 &=& \frac{(M_J^2-s_1)^2 \, (M_V^2-s_1)^2}
      {(M_{f_i}^2-s_1)^2 +M_{f_i}^2 \Gamma_{f_i}^2} 
+ \frac{(M_J^2-s_3)^2 \, (M_V^2-s_3)^2}
  {(M_{f_i}^2-s_3)^2 +M_{f_i}^2 \Gamma_{f_i}^2} \nonumber\\
&+& \frac{(M_{f_i}^2-s_1)(M_{f_i}^2-s_3) + M_{f_i}^2 \Gamma_{f_i}^2} 
{((M_{f_i}^2-s_1)^2 +M_{f_i}^2 \Gamma_{f_i}^2)
 ((M_{f_i}^2-s_3)^2 +M_{f_i}^2 \Gamma_{f_i}^2)} \, 
\biggl((s_1s_3 - M_J^2 M_V^2)^2 + s_2^2 M_J^2 M_V^2\biggr)\,, \\
R_3 &=& \frac{s_2^2}{(M_{f_i}^2 - s_2)^2 + M_{f_i}^2 \Gamma_{f_i}^2} \ 
\biggl( 
 \frac{(M_{f_i}^2-s_1)(M_{f_i}^2-s_2) + M_{f_i}^2 \Gamma_{f_i}^2}
{(M_{f_i}^2-s_1)^2 + M_{f_i}^2 \Gamma_{f_i}^2} \, 
 (s_1^2 + M_J^2 M_V^2) \nonumber\\
 &+& \frac{(M_{f_i}^2-s_3)(M_{f_i}^2-s_2) + M_{f_i}^2 \Gamma_{f_i}^2} 
{(M_{f_i}^2-s_3)^2 + M_{f_i}^2 \Gamma_{f_i}^2} \, 
(s_3^2 + M_J^2 M_V^2) \biggr) \, . 
\en 

The decay width $\Gamma_{\rm III} \equiv 
\Gamma(\Jpsi \to f_i \gamma \to \gamma P P)$  
is given by: 
\eq 
\hspace*{-.5cm}
\Gamma_{\rm III} = \frac{N_{PP}}{768 \pi^3 M_J^3}  
\int\limits_{4M_P^2}^{M_J^2}
ds_2 \int\limits_{s_{\rm 1, III}^-}^{s_{\rm 1, III}^+}ds_1 
\sum_{\rm pol}|M_{\rm III}|^2  
= \frac{\alpha \, g_{Jf_i\gamma}^2 \, g_{f_iPP}^2}
{384 \pi^2 \,  M_J^3} \, N_{PP} 
\int\limits_{4M_P^2}^{M_J^2} ds_2 
\ \sqrt{1-\frac{4M_P^2}{s_2}}  
\frac{(M_J^2  - s_2)^3}{(M_{f_i}^2 - s_2)^2 
+ M_{f_i}^2 \Gamma_{f_i}^2} \,, 
\en 
where  
\eq
s_{\rm 1, III}^{\pm} = M_P^2 + \frac{M_J^2-s_2}{2} 
\biggl( 1 \pm \sqrt{1-\frac{4M_P^2}{s_2}} \ \biggr) \,.  
\en 
Using Eqs.~(\ref{gfPP}) and (\ref{gJfg}) we again express $\Gamma_{\rm III}$ 
through the two-body decay widths $\Gamma(\Jpsi \to f_i \gamma)$ and 
$\Gamma(f_i \to PP)$ as: 
\eq\label{GammaIII_fin} 
\Gamma_{\rm III} &=& \frac{\Gamma(\Jpsi \to f_i \gamma) \, 
\Gamma(f_i \to PP) \, M_{f_i}^2}{\pi \, 
(M_J^2 - M_{f_i}^2)^3 \, \sqrt{M_{f_i}^2 - 4M_P^2}} \ 
\int\limits_{4M_P^2}^{M_J^2} ds_2 
\ \sqrt{1-\frac{4M_P^2}{s_2}} \, 
\frac{(M_J^2  - s_2)^3}{(M_{f_i}^2 - s_2)^2 
+ M_{f_i}^2 \Gamma_{f_i}^2} \,, 
\en 

The decay width $\Gamma_{\rm IV} \equiv  
\Gamma(J/\psi \to f_i V(\gamma)  \to \gamma V V)$ 
is determined according to: 
\eq 
\label{GammaIV}
\Gamma_{\rm IV} = \frac{1}{1536 \pi^3 M_J^3} 
\int\limits_{4M_V^2}^{M_J^2}ds_2 
\int\limits_{s_{\rm 1, IV}^-}^{s_{\rm 1, IV}^+} 
ds_1 \sum_{\rm pol}|M_{\rm IV}|^2 
\en 
where 
\eq 
s_{\rm 1, IV}^{\pm} = M_V^2 + \frac{M_J^2-s_2}{2} 
\biggl( 1 \pm \sqrt{1-\frac{4M_V^2}{s_2}} \ \biggr) \, . 
\en
Again, the sum can be expressed as
\eq
\sum_{\rm pol}|M_{\rm IV}|^2 
= \pi \alpha \biggl( g_{Jf_i\gamma}^2 \ g_{f_iVV}^2 \ Q_1 
 + g_{Jf_iV}^2 \ g_{f_iV\gamma}^2 \ Q_2
 + g_{Jf_iV} \ g_{Jf_i\gamma} \ g_{f_iVV} \ g_{f_iV\gamma} 
\ Q_3 \biggr) 
\en 
with
\eq 
Q_1 &=& \frac{1}{(M_{f_i}^2 - s_2)^2 + M_{f_i}^2 \Gamma_{f_i}^2} \ 
\biggl( s_2^4 - 2 s_2^3 (M_J^2 + 2M_V^2) + s_2^2 
(M_J^4 + 8 M_J^2 M_V^2 + 6 M_V^4) \nonumber\\ 
&-& 4 s_2 M_J^2 M_V^2 (M_J^2 + 3 M_V^2)
+ 6 M_J^4 M_V^4 \biggr) \,, \\ 
Q_2 &=& \frac{1}{(M_{f_i}^2-s_1)^2 +M_{f_i}^2 \Gamma_{f_i}^2} 
\biggl( s_1^4 - 2 s_1^3 (M_J^2 + 2 M_V^2) + 
s_1^2 (M_J^4 + 8 M_J^2 M_V^2 + 6 M_V^4)\nonumber\\  
&-& 2 s_1 M_V^2 ( M_J^4 + 5M_J^2 M_V^2 + 2 M_V^4) 
+ M_V^4 (M_J^4 + 4 M_J^2 M_V^2 + M_V^4) \biggr) \nonumber\\
&+&\frac{1}{(M_{f_i}^2-s_3)^2 + M_{f_i}^2 \Gamma_{f_i}^2} 
\biggl( s_3^4 - 2 s_3^3 (M_J^2 + 2 M_V^2) 
+ s_3^2 (M_J^4 + 8 M_J^2 M_V^2 + 6 M_V^4)\nonumber\\  
&-& 2 s_3 M_V^2 ( M_J^4 + 5M_J^2 M_V^2 + 2 M_V^4) 
+ M_V^4 (M_J^4 + 4 M_J^2 M_V^2 + M_V^4) \biggr) \nonumber\\
&+&\frac{(M_{f_i}^2-s_1) \, (M_{f_i}^2-s_3) + M_{f_i}^2 \Gamma_{f_i}^2} 
{((M_{f_i}^2-s_1)^2 + M_{f_i}^2 \Gamma_{f_i}^2) \, 
((M_{f_i}^2-s_3)^2 + M_{f_i}^2 \Gamma_{f_i}^2)} \, 
\biggl( s_1^2s_3^2 + (s_1^2 + s_3^2) M_V^2 (M_J^2 + M_V^2)\nonumber\\  
&-& 2 (s_1 + s_3) M_V^4 (M_J^2 + 2M_V^2) + 2 M_V^6 (2M_J^2 + 5M_V^2) \biggr)
\,, \\
Q_3 &=& \frac{1}{(M_{f_i}^2 - s_2)^2 + M_{f_i}^2 \Gamma_{f_i}^2} \ 
\biggl( 
 \frac{(M_{f_i}^2-s_1)(M_{f_i}^2-s_2) + M_{f_i}^2 \Gamma_{f_i}^2} 
{(M_{f_i}^2-s_1)^2 + M_{f_i}^2 \Gamma_{f_i}^2} \, 
 \biggl (s_1^2s_2^2 + 2s_1^2 M_J^2 M_V^2 
+ s_2^2 M_V^2 (M_J^2 + M_V^2) \nonumber\\ 
&-& 2 (s_1 + s_2) M_J^2 M_V^2 (M_J^2 + 2 M_V^2) 
+ M_J^2 M_V^2 ( M_J^4 + 4 M_J^2 M_V^2 + 2 M_V^4 ) \biggr)\nonumber\\ 
&+& 
 \frac{(M_{f_i}^2-s_3)(M_{f_i}^2-s_2) + M_{f_i}^2 \Gamma_{f_i}^2}
{(M_{f_i}^2-s_3)^2 + M_{f_i}^2 \Gamma_{f_i}^2} \, 
 \biggl(s_3^2s_2^2 + 2s_3^2 M_J^2 M_V^2 
+ s_2^2 M_V^2 (M_J^2 + M_V^2) \nonumber\\ 
&-& 2 (s_3 + s_2) M_J^2 M_V^2 (M_J^2 + 2 M_V^2) 
+ M_J^2 M_V^2 ( M_J^4 + 4 M_J^2 M_V^2 + 2 M_V^4 ) \biggr) \biggr) \, .
\en

\newpage 

  \begin{figure} 
  \centering{\
  \epsfig{figure=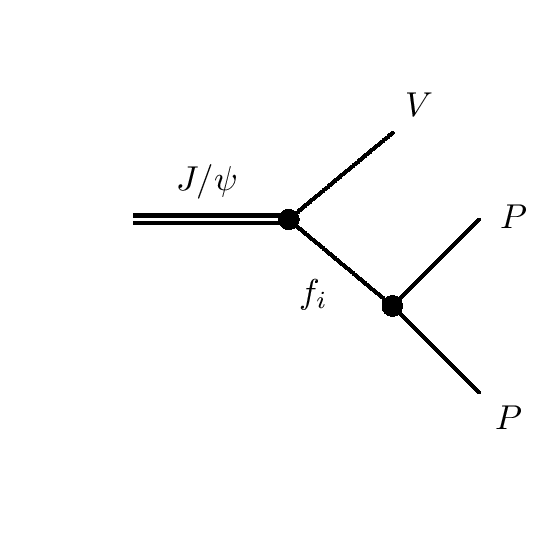,scale=.75}}
  \caption{
  Contribution of the scalar resonances 
  $f_i = f_0(1370)$, $f_0(1500)$,  
  $f_0(1710)$ to the decay 
  $J/\psi \to f_i V \to V P P$}  

  \centering{\
  \epsfig{figure=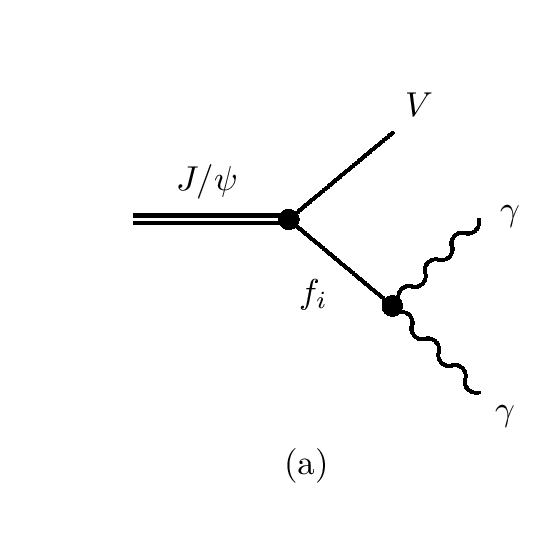,scale=.75}
  \epsfig{figure=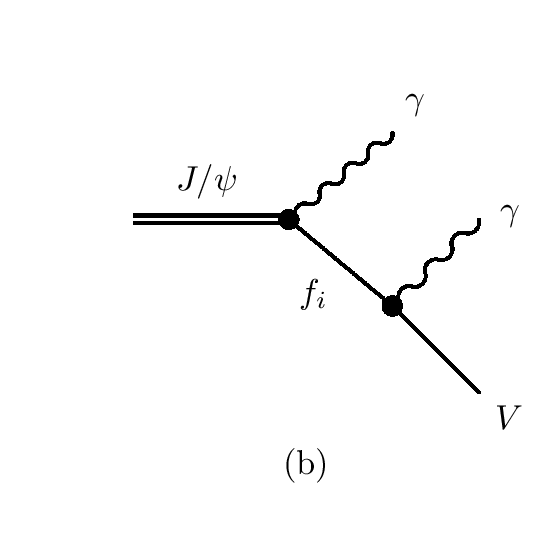,scale=.75}}
  \caption{Contribution of the scalar resonances 
  $f_i = f_0(1370)$, $f_0(1500)$,  
  $f_0(1710)$ to the decay 
  $J/\psi \to f_i V(\gamma) \to V \gamma \gamma $}  

  \centering{\
  \epsfig{figure=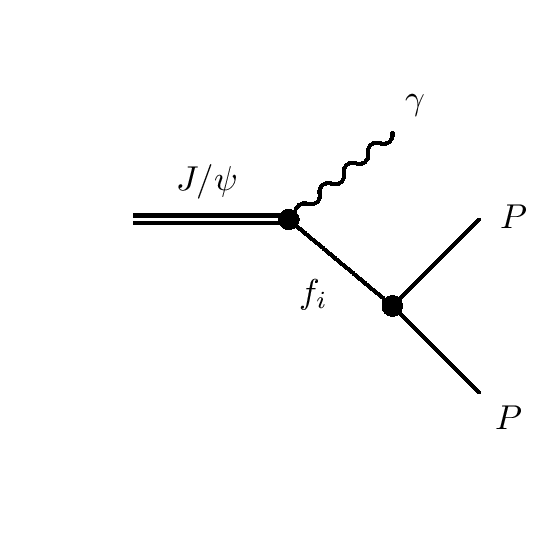,scale=.75}}
  \caption{Contribution of the scalar resonances 
  $f_i = f_0(1370)$, $f_0(1500)$,  
  $f_0(1710)$ to the decay  
  $J/\psi \to f_i \gamma \to \gamma P P $}  
  \centering{\
  \epsfig{figure=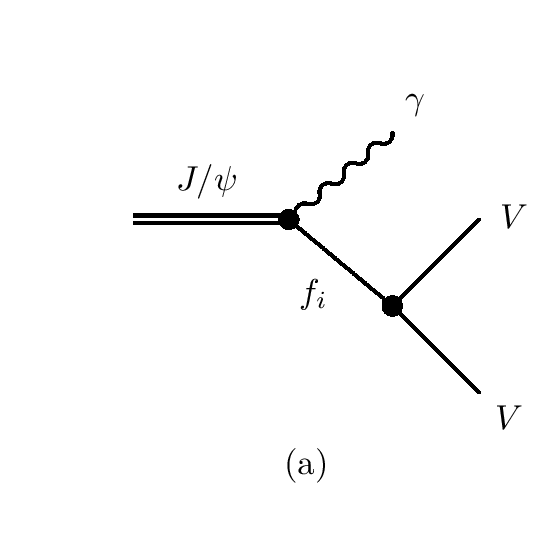,scale=.75}
  \epsfig{figure=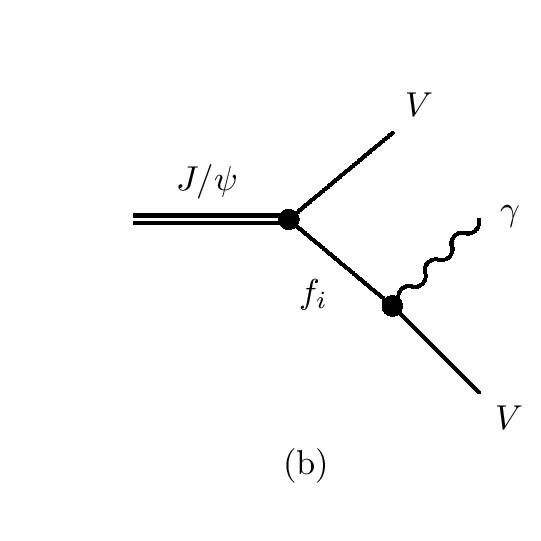,scale=.75}}
  \caption{Contribution of the scalar resonances 
  $f_i = f_0(1370)$, $f_0(1500)$,  
  $f_0(1710)$ to the decay 
  $J/\psi \to f_i V(\gamma) \to \gamma V V $}  
  \end{figure}

\newpage 

\begin{table} 
\caption{Masses and decay properties of the scalar mesons $f_i$.}

  \centering
  \begin{tabular}{|l|r@{ $\pm$ }l|r|r|r|r|}
    \hline
    \multicolumn{1}{|l|}{Quantity}       &
    \multicolumn{2}{c|}{Data~\cite{PDG}} &
    \multicolumn{1}{c|}{Fit I}           &
    \multicolumn{1}{r|}{$\chi^2$}        &
    \multicolumn{1}{c|}{Fit II}          &
    \multicolumn{1}{r|}{$\chi^2$}\\
    \hline
    $M_{f_1}$  (MeV) &  1350&   200&  1432&     0.16&  1231&     0.36\\
    $M_{f_2}$  (MeV) &  1505&     6&  1510&     0.72&  1510&     0.70\\
    $M_{f_3}$  (MeV) &  1720&     6&  1720& $\sim 0$&  1720& $\sim 0$\\
    $\Gamma_{f_2\to\eta\eta'}$  (MeV) 
&  2.07&  0.87&  1.2&     0.97&  1.02&     1.45\\
    $\Gamma_{f_2\to\eta\eta}$ (MeV)   
&  5.56&  0.98&  2.8&     7.98&  4.70&     0.76\\
    $\Gamma_{f_2\to\bar{K}K}$ (MeV)   
&  9.37&  1.09& 10.4 &  0.96 & 8.5&     0.19\\
    $\Gamma_{f_2\to\pi\pi}$   (MeV)   
& 38.04&  2.51& 37.7 &  0.02& 38.38&     0.02\\
    $\Gamma_{f_3\to\eta\eta}/\Gamma_{f_3\to\bar{K}K}$ 
&  0.48&  0.15&  0.25&     2.45&  0.25&     2.45\\
    $\Gamma_{f_3\to\pi\pi}/\Gamma_{f_3\to\bar{K}K}$  
&  0.41&  0.14&  0.43&     0.03&  0.49&     0.36\\
    $\Gamma_{f_3}\to{2P}$     (MeV)   
&   137&     8&   137& $\sim 0$&   136&     0.01\\
    $\Gamma_{a_0\to\pi\eta'}/\Gamma_{a_0\to\pi\eta}$ 
&  0.35&  0.16& 0.29&    0.15&  0.29&     0.13\\
    $\Gamma_{a_0\to KK'}/\Gamma_{a_0\to\pi\eta}$     
&  0.88&  0.23&  0.8&    0.11&  1.06&     0.64\\
    $\Gamma_{K_0^*(1430)\to K\pi}$  (MeV)            
&   251& 74.40& 64.7&    6.27& 41.64&     7.92\\
    \hline
    \multicolumn{1}{|l}{$\chi^2_{\rm tot}$}&
    \multicolumn{3}{l}{}              &
    \multicolumn{1}{|r|}{19.82}      &
    \multicolumn{1}{l}{}              &
    \multicolumn{1}{|r|}{14.99}\\
    \hline
  \end{tabular}

  \begin{center}
    \caption{Comparison of strong scalar decays  with WA102 data~\cite{WA102}.} 

    \begin{small}
    \begin{tabular}{|c|c|c|c|}
      \hline
      Quantity & WA102 data~\cite{WA102} & I & II \\
      \hline
      $\Gamma_{f_1\to\bar{K}K}/\Gamma_{f_1\to\pi\pi}$ 
& 0.46 $\pm$ 0.19 & 1.07 & 2.27 \\ 
      $\Gamma_{f_1\to\eta\eta}/\Gamma_{f_1\to\pi\pi}$ 
& 0.16 $\pm$ 0.07 & 0.22 & 0.4 \\ 
      $\Gamma_{f_1\to 2P}$ (MeV)  
                         & small    & 314 & 131 \\ 
      $\Gamma_{f_3\to\eta\eta'}/\Gamma_{f_3\to\pi\pi}$ 
& $< 0.18$        & 0.89 & 1.4 \\ 
      \hline
    \end{tabular} 
  \end{small}
\end{center}


  \begin{center}
    \caption{Branchings of the $\Jpsi$ hadronic decays in units $10^{-4}$.}
    
    \label{TableJpsiHadronic}
    \begin{small}
    \begin{tabular}{|l |c|c|c| c | c|c|c| c |c|c|c| c |c|c|c| c|}    
      \hline
      Meson & 
      \multicolumn{3}{c|}{$\phi K\bar{K}$}   & 
      Data~\cite{PDG}  &  
      \multicolumn{3}{c|}{$\omega K\bar{K}$} & 
      Data~\cite{PDG}  & 
      \multicolumn{3}{c|}{$\phi\pi\pi$}      & 
      Data~\cite{PDG}  & 
      \multicolumn{3}{c|}{$\omega\pi\pi$}    & 
      Data~\cite{PDG} \\
      \hline 
                 & I & II & \cite{Close:2005vf} &              
                 & I & II & \cite{Close:2005vf} &                
                 & I & II & \cite{Close:2005vf} &                            
                 & I & II & \cite{Close:2005vf} & \\ 
      $f_3$ & 3.6  & 2.5 & 3.6  &  $3.6\pm 0.4$   & 
              4.8  & 5.2 & 4.8  &  $4.8\pm 1.1$ & 
              1.7  & 1.3 & 0.40 & -               &
              2.2  & 2.7 & 0.53 & - \\ 

      $f_2$ & 0.5  & 2.3 & 0.19 & $0.8\pm 0.5$    & 
              0.2  & 1.1 & 1.38 & -               & 
              2.0  & 9.4 & 0.77 & $1.7\pm 0.8$    & 
              0.6  & 4.6 & 5.60 & - \\ 

      $f_1$ & 4.1  & 4.7 & 0 & $0.3\pm 0.3$       &
              1.3  & 8.7 & 0.08 & -               & 
              4.3  & 2.4 & 0.01 & $4.3\pm 1.1$    & 
              1.4  & 4.4 & 0.83 & - \\ 
      \hline
    \end{tabular}
    \end{small}
  \end{center} 
\end{table} 

\newpage 

\begin{table} 

  \begin{center}
    \caption{The branchings of the $\Jpsi$ two-body decays in $10^{-4}$ and 
    effective couplings in $10^{-3}$ $\times$ GeV$^{-1}$.} 

    \begin{small}
    \begin{tabular}{|l|c|c|c|}
      \hline
      Quantity & I & II & Data~\cite{PDG} \\      
      \hline
      $g_{Jf_1\omega}$ & $-$ 0.60 & $-$ 1.91 &\\ 
      $g_{Jf_2\omega}$ & 0.57 & 2.2 & \\ 
      $g_{Jf_3\omega}$ & $-$ 1.60 & $-$ 1.80& \\ 
      $g_{Jf_1\phi}$   & 1.02 & 1.35 & \\ 
      $g_{Jf_2\phi}$   & -1.00 & 2.18 &\\ 
      $g_{Jf_3\phi}$   & -1.40 & 1.27 & \\ 
      $g_{Jf_1\gamma}$ & 2.34 & 2.12 &\\ 
      $g_{Jf_2\gamma}$ & $-$ 1.53 & $-$ 1.53& \\ 
      $g_{Jf_3\gamma}$ & $-$ 7.42 & 7.98 &\\  
      ${\rm Br}(\Jpsi\to f_1\omega)$ & 2.2 & 27.5 &\\ 
      ${\rm Br}(\Jpsi\to f_2\omega)$ & 1.8 & 13.6 &\\ 
      ${\rm Br}(\Jpsi\to f_3\omega)$ & 10.7 & 13.5 &\\ 
      ${\rm Br}(\Jpsi\to f_1\phi)$   & 6.8  & 14.9 &\\ 
      ${\rm Br}(\Jpsi\to f_2\phi)$   & 5.9  & 28.0 &\\ 
      ${\rm Br}(\Jpsi\to f_3\phi)$   & 8.1 & 6.6 &\\ 
      ${\rm Br}(\Jpsi\to f_1\gamma)$ & 2.6 & 2.6 &\\
      ${\rm Br}(\Jpsi\to f_2\gamma)$ & 1.0 & 1.0 & 1.01 $\pm$ 0.32 \\
      ${\rm Br}(\Jpsi\to f_3\gamma)$ & 17.7 & 20.5 &\\
      \hline
    \end{tabular} 
  \end{small}
\end{center}

  \begin{center}
    \caption{Effective couplings of $f_i$ in units GeV$^{-1}$. 
}
 
    \begin{tabular}{|l|c|c|}
      \hline
      Coupling & I & II  \\
      \hline
$g_{f_1\gamma\gamma}$                         & 0.30       &  0.41\\ 
$g_{f_2\gamma\gamma}$                         & $-$ 0.21   &  0.13\\ 
$g_{f_3\gamma\gamma}$                         & $-$ 0.01   &  $-$ 0.08\\ 
$g_{f_1 \rho\gamma} = 3 g_{f_1 \omega\gamma}$ & 1.24 & 1.60 \\
$g_{f_1 \phi\gamma}$                          & 0.29 & 0.75 \\
$g_{f_2 \rho\gamma} = 3 g_{f_2 \omega\gamma}$ & $-$ 0.90 & 0.98 \\
$g_{f_2 \phi\gamma}$                          & $-$ 0.13 & $-$ 0.94 \\
$g_{f_3 \rho\gamma} = 3 g_{f_3 \omega\gamma}$ & $-$ 0.47 & $-$ 0.44 \\
$g_{f_3 \phi\gamma}$                          & 0.10    &  $-$ 0.20 \\
$g_{f_1\rho\rho} = g_{f_1\omega\omega}$       & 6.19    &  8.01 \\ 
$g_{f_1 \phi\phi}$                            & 3.09    &  8.00 \\ 
$g_{f_2\rho\rho} = g_{f_2\omega\omega}$       & $-$ 4.44   &  4.91  \\ 
$g_{f_2 \phi\phi}$                            & $-$ 1.33   &  $-$ 9.97 \\ 
$g_{f_3\rho\rho} = g_{f_3\omega\omega}$       & $-$ 2.35   &  $-$ 2.19 \\ 
$g_{f_3 \phi\phi}$                            & 10.74  &   $-$ 2.16 \\ 
\hline  
\end{tabular} 
\end{center}
\end{table}  

\newpage 
\begin{table} \label{TableJpsiRad1}
  \begin{center}
    \caption{Branchings of the $\Jpsi$ radiative decays in 
    units $10^{-4}$.}    

    \begin{tabular}{|l|c|c|c|}
      \hline
      Mode & I & II & Data~\cite{PDG} \\      \hline
      $\Jpsi\to f_1\gamma\to \gamma \pi\pi$    & 1.65 & 0.42 & \\
      $\Jpsi\to f_2\gamma\to \gamma \pi\pi$    & 0.34 & 0.34 & \\
      $\Jpsi\to f_3\gamma\to \gamma \pi\pi$    & 4.48 & 5.02 & 4.0 $\pm$ 1.0\\
      $\Jpsi\to f_1\gamma\to \gamma K\bar{K}$  & 1.58 & 0.82 & \\
      $\Jpsi\to f_2\gamma\to \gamma K\bar{K}$  & 0.09 & 0.08 & \\
      $\Jpsi\to f_3\gamma\to \gamma K\bar{K}$  & 8.00 & 7.85 
      & $8.5^{+1.2}_{-0.9}$ \\
      $\Jpsi\to f_1\gamma\to \gamma \omega\omega$ & 0.33 & 0.22 & \\
      $\Jpsi\to f_2\gamma\to \gamma \omega\omega$ & 0.13 & 0.16 & \\
      $\Jpsi\to f_3\gamma\to \gamma \omega\omega$ & 3.09 & 3.07 & 3.1 $\pm$ 1.0\\
      $\Jpsi\to f_1\gamma\to\rho\gamma\gamma$     
      & 2.0 $\times 10^{-2}$&  1.7 $\times 10^{-2}$ & \\
      $\Jpsi\to f_1\omega(\gamma)\to\omega\gamma\gamma$ 
      & 2.2 $\times 10^{-3}$&  3.3 $\times 10^{-3}$ & \\
      $\Jpsi\to f_1\phi(\gamma)\to\phi\gamma\gamma$     
      & 0.8 $\times 10^{-3}$&  2.5 $\times 10^{-3}$ & \\
      $\Jpsi\to f_2\gamma\to\rho\gamma\gamma$     
      & 0.9 $\times 10^{-2}$&  1.1 $\times 10^{-2}$ & \\
      $\Jpsi\to f_2\omega(\gamma)\to\omega\gamma\gamma$ 
      & 1.1 $\times 10^{-3}$&  1.5 $\times 10^{-3}$ & \\
      $\Jpsi\to f_2\phi(\gamma)\to\phi\gamma\gamma$     
      & 0.4 $\times 10^{-4}$&  0.5 $\times 10^{-2}$ & \\
      $\Jpsi\to f_3\gamma\to\rho\gamma\gamma$     
      & 6.0 $\times 10^{-2}$&  6.5 $\times 10^{-2}$ & \\
      $\Jpsi\to f_3\omega(\gamma)\to\omega\gamma\gamma$ 
      & 0.7 $\times 10^{-2}$ & 0.7 $\times 10^{-2}$ & \\
      $\Jpsi\to f_3\phi(\gamma)\to\phi\gamma\gamma$     
      & 0.16 &  0.8 $\times 10^{-2}$ & \\ 
    \hline
    \end{tabular}
  \end{center} 

  \caption{Decay widths of $f_i\to V\gamma$ and $f_i\to \gamma\gamma$ 
transitions (in keV) in comparison with other theoretical predictions.} 

  \centering
  \begin{tabular}{|c|c|c|c|c|c|c|}
\hline
 Mode & Ref.~\cite{NRQM}& Ref.~\cite{LFQM}& Ref.~\cite{Nagahiro:2008bn}    
      & Ref.~\cite{Branz:2009cv} &  Ref.~\cite{Giacosa:2005qr} 
& Our (I, II) in keV\\
\hline
$f_1\to\gamma\gamma$ & & $(1.6, 3.9_{-0.7}^{+0.8}, 5.6_{-1.3}^{+1.4})$  
& & 0.35 & 1.31 & (11.1, 13.4) \\ 
$f_2\to\gamma\gamma$ & & & & 0.35 & & (6.5, 2.4)\\ 
$f_3\to\gamma\gamma$ & & $(0.92, 1.3_{-0.2}^{+0.2}, 3.0_{-1.2}^{+1.4})$ 
& & 0.019 & 0.05 & (0.02, 1.2)\\ 
$f_1\to\rho\gamma$   & $(443,1121,1540)$ & $(150, 390_{-70}^{+80}, 530_{-110}^{+120})$ & $(79 \pm 40, 125 \pm 80)$ & & 726 & (1441, 957)\\
$f_1\to\omega\gamma$ & & & $(7 \pm 3, 128 \pm 80)$ & & 0.04 & (156, 102)\\
$f_1\to\phi\gamma$   & $(8,9,32)$ & $(0.98, 0.83_{-0.23}^{+0.27}, 
4.5_{-3.0}^{+4.5})$ & $(11 \pm 6, - )$ & & 0.01 & (27, 30)\\
$f_2\to\rho\gamma$   & & & & & & (990, 1209)\\
$f_2\to\omega\gamma$ & & & & & & (108, 131)\\
$f_2\to\phi\gamma$   & & & & & & (8, 447)\\
$f_3\to\rho\gamma$   & $(42,94,705)$ & $(24, 55_{-14}^{+16}, 410_{-160}^{+200})$ & $(100 \pm 40, - )$ & & 24 & (519, 450)\\
$f_3\to\omega\gamma$ & & & $(3.3 \pm 1.2, - )$ & & 82 & (57, 49)\\
$f_3\to\phi\gamma$   & $(800,718,78)$ & $(450, 400_{-20}^{+20}, 36_{-14}^{+17})$ & $(15 \pm 5, - )$ & & 94 & (1300, 53)\\
\hline
\end{tabular}
\end{table}

\end{document}